\documentclass[conference]{IEEEtran}
\makeatletter

\usepackage{blindtext}
\usepackage{cite}
\usepackage{amsmath,amssymb,amsfonts}
\usepackage{algorithmic}
\usepackage{graphicx}
\usepackage{textcomp}
\usepackage{orcidlink}
\usepackage{listings}
\usepackage{tcolorbox}
\usepackage{dirtytalk}
\usepackage{multirow}
\usepackage{tabularray}
\usepackage{eso-pic}
\usepackage{xcolor}
\usepackage{tcolorbox}
\usepackage{tabularx}
\usepackage{colortbl}
\def\BibTeX{{\rm B\kern-.05em{\sc i\kern-.025em b}\kern-.08em
    T\kern-.1667em\lower.7ex\hbox{E}\kern-.125emX}}

\newcommand\llm{Large Language Model}
\newcommand\llms{Large Language Models}
\newcommand\llmabr{LLM}
\newcommand\llmsabr{LLMs}
\newcommand\ide{IDE}
\newcommand\ai{AI}
\newcommand\apps{\ai{} Pair Programmers}
\newcommand\app{\ai{} Pair Programmer}

\newcommand\totalSurveyPapers{23}

\newcommand\categoryOne{Source of Assistance}
\newcommand\categoryTwo{Assistance in SDLC Steps}
\newcommand\categoryThree{Source Code Assistance Scope}
\newcommand\categoryFour{Software Development Domain}
\newcommand\categoryFive{Knowledge Level of Developers}
\newcommand\categorySix{Level of Collaboration}
\newcommand\categorySeven{Understanding Mental Model}
\newcommand\categoryEight{Impacts in Learning }
\newcommand\categoryNine{Benefits}
\newcommand\categoryTen{Concerns}
\newcommand\categoryEleven{Cognitive Factors}

\begin{document}

\title{\vspace*{1cm} The Transformative Influence of LLMs on Software Development \& Developer Productivity\\
}

\author{\IEEEauthorblockN{Sajed Jalil \orcidlink{0000-0003-1249-2113}}
\IEEEauthorblockA{
\textit{George Mason University}\\
Virginia, USA \\
sjalil@gmu.edu}
}


\maketitle

\begin{abstract}
The increasing adoption and commercialization of generalized \llms{} (\llmsabr{}) have profoundly impacted various aspects of our daily lives. Initially embraced by the computer science community, the versatility of \llmsabr{} has found its way into diverse domains. In particular, the software engineering realm has seen the most transformative changes. With \llmsabr{} increasingly serving as \ai{} Pair Programming Assistants spurred the development of specialized models aimed at aiding software engineers. Although this new paradigm offers numerous advantages, it presents critical challenges and open problems.

To identify the potential and prevailing obstacles, we systematically reviewed contemporary scholarly publications, emphasizing the perspectives of software developers and usability concerns. Preliminary findings underscore pressing concerns about data privacy, bias, and misinformation. Additionally, we identified several usability challenges, including prompt engineering, increased cognitive demands, and mistrust. Finally, we introduce 12 open problems identified through our survey, covering these domains.

\end{abstract}


\section{Introduction}
Just a few years ago, software developers relied largely on the rudimentary code suggestions and auto-completions offered by Interactive Development Environments (\ide{}). The support rendered by these \ide{}s was often confined to specific programming languages, domains, and platforms~\cite{sorte2015use}. When confronted with a task hurdle, developers commonly turned to Q/A forums such as Stack Overflow~\cite{sof}, consulted existing documentation, or sought advice from colleagues~\cite{liu2023better}. This reliance often demanded considerable time and effort, in addition to incurring heavy costs during context switching. Moreover, prior research indicates that platforms like Stack Overflow are not particularly welcoming for novices~\cite{al2022empirical,calefato2015mining}. Such a setting often discourages beginners from asking questions, prompting them to browse through the existing answers on similar issues. Often, one must wait a considerable amount of time before someone knowledgeable is available to help.

To improve the capability of \ide{}s, various third-party plugins, extensions, and tool support began to appear~\cite{murphy2006java, leich2005tool}. These add-ons offer developers a myriad of functionalities, ranging from advanced debugging capabilities to vulnerability detection, automatic compilation, and optimization. However, developers often skipped using these tools because it was not readily apparent how to use them to get the best result. This is confirmed by Beller \textit{et al.} in a study that finds that the developer often avoids using debugging tools due to difficulty and perceives console prints as the most convenient form of debugging~\cite{beller2018dichotomy}. Furthermore, as \ide{}s continue to incorporate an increasing number of these tools, they began to clutter the interface, making \ide{} more distracting and cognitively demanding.

Over time, the rise of machine learning and deep learning plugins, coupled with enhanced usability features, became evident~\cite{cito2018performancehat, mamede2022transformer}. With the increasing capability and availability of better Graphical Processing Units (GPUs), toolmakers began to train these \ai{} models with more and more data. An early example, Code2vec\cite{alon2019code2vec}, was trained on 12 million methods sourced from GitHub\cite{github}. However, translating natural language into programming language posed challenges. Unlike natural languages, the potential terms or variable names in programming languages are virtually limitless. Subsequent models, such as CodeBert~\cite{feng2020codebert} and Code2Seq~\cite{alon2018code2seq}, tackled this challenge and introduced further refinements. This race continued and paved the way for the emergence of massive \llms{} with estimated trillions of parameters such as ChatGPT~\cite{chatgpt}, Claude~\cite{claude}, Perplexity~\cite{perplexity}.

With the emergence and subsequent popularity of \llmsabr{}, there has been a noticeable transformation in the daily activities of software developers~\cite{sofSurvey}. \llmsabr{}, can expedite numerous software development tasks, making them more straightforward. They are equipped to handle a broad spectrum of tasks via natural language, code input, or a combination of both. Several specialized tools, such as GitHub Copilot\cite{copilot}, Cursor\cite{cursor}, Devin\cite{devin}, and Tabnine~\cite{tabnine}, have been crafted to specifically address software development needs. They leverage vast software artifacts available across the web for their training along with various agentic workflows.

These specialized \llmsabr{} emulate the dynamics of human pair programming and are often referred to as \apps{}. Even though they come with both benefits and limitations, like human collaborators, their pros and cons are distinct because they are not human entities. To ensure that they align well with the thought processes of software developers, rigorous research is essential. As such, it is crucial to grasp the elements influencing these models and identify any usability challenges they present.

In this paper, we present a detailed survey of scholarly articles, delineated in Table~\ref{tab:publications_list} and Table~\ref{tab:reports_list}. We aim to foster a substantive contribution to the realm of Software Engineering (SE) and \apps{}. We aim to make the following contributions:
\begin{enumerate}
    \item \textbf{Examining the evolutionary trajectories in software engineering in response to AI tools adoption}.
    \item \textbf{A classification of factors steering these trends}.
    \item \textbf{Evaluating what is working and what is not working with this approach}.
    \item \textbf{Identifying the existing open problems}.
\end{enumerate}

\begin{table*}
    \centering
    \caption{Statistics of the Selected Publications}
    \label{tab:publications_list}
    
    \begin{tblr}{colspec={|r|c|c|c|c|c|c|r|c|c|},
        vlines, hlines, vspan=even,
        row{1} = {bg=blue, fg=white},
        rowsep=0.2pt}
          S.N. & Conf. Code & Year & Citations & Tools & {Industry\\Association} & {User\\Study} & {Participant\\Count} & {Dataset\\Evaluation} & Ref. \\\hline

          1&EASE&  2023&  172&  ChatGPT&  &  Y&  1&  & \cite{ahmad2023towards} \\ 
          2&arXiv&  2022&  133&  Copilot, Tabnine&  Y&  &  &  Human Eval\cite{humaneval}& \cite{sarkar2022like} \\
          3&MSR&  2022&  370&  Copilot&  &  &  &  & \cite{nguyen2022empirical} \\ 
          4&ICSE&  2022&  176&  Copilot&  &  Y&  21&  & \cite{imai2022github} \\ 
          5&PROMISE&  2022&  164&  Copilot&  &  &  &  Human Eval\cite{humaneval}& \cite{yetistiren2022assessing} \\ 
          6&IEEE Software&  2023&  261&  Copilot, ChatGPT,  Tabnine, Others &  Y&  Y&  -&  & \cite{ebert2023generative} \\ 
          7&IJAEGT&  2015&  39&  &  &  &  &  & \cite{sorte2015use} \\ 
          8&IJHS&  2016&  322&  Siri, Cortana, Google Now&  &  &  &  & \cite{canbek2016track}  \\ 
          9&ACM Tran.&  2022&  87&  Themisto&  &  Y&  24&  & \cite{wang2022documentation}  \\ 
         10& CHI& 2023& 93& Copilot, Tabnine, Others& Y& Y& 15& & \cite{mcnutt2023design}\\
         11& ICSE& 2024& 167& Copilot, Tabnine& & Y& 410& & \cite{liang2023understanding}\\
         12& CHI& 2022& 93& & & & & & \cite{zheng2022ux}\\
         13& ICSE& 2023& 29& VS IntelliCode& Y& Y& 61& & \cite{vaithilingam2023towards}\\
         14& CHI& 2023& 131& Excel, Codex& Y& Y& 24& Stack Overflow~\cite{stackooverflowexcel}&\cite{liu2023wants}\\
         15& CACM & 2022& 35& Copilot& Y& Y& 2047& &\cite{bird2023taking}\\
         16& arXiv& 2023& 161& ChatGPT& & Y& -& &\cite{yuan2023no}\\
         17& ICSE& 2022& 245& GPT-3, Codex& Y& Y& -& &\cite{jain2022jigsaw}\\
         18& arXiv& 2023& 53& Copilot& Y& Y& 17& &\cite{wang2023investigating}\\
         19& Systems & 2023& 29& AeroBERT-NER \& Classifier, Flair & Y& & & NER dataset~\cite{tikayat2023aerobert}&\cite{tikayat2023agile}\\
         20& PACMPL & 2023& 403& Copilot & & Y& 20& &\cite{barke2023grounded}\\
         
    \end{tblr}
\end{table*}

\begin{table*}
    \centering
    \caption{Statistics of the Selected Surveys, Blogs \& Articles}
    \label{tab:reports_list}
    \begin{tblr}{ colspec={|c|l|c|c|c|c|},
        vlines, hlines, vspan=even,
        row{1} = {bg=blue, fg=white},
        rowsep=0.2pt}
         S.N. &Name&  Type&  Year&  Industry Association&  Ref.\\ \hline
         1 &Stack Overflow Developer Survey&  Survey&  2024&  Y&  \cite{sofSurvey}\\
         2 & Real-World Programming with ChatGPT & Article & 2023&  &\cite{oreilly}\\  
         3 & How to build an enterprise LLM application: Lessons from GitHub Copilot&  Blog&  2023&  Y& \cite{githubTutorial}\\ 
    \end{tblr}
\end{table*}

In the following sections, we will offer an in-depth analysis of our survey to address these inquiries. To begin, \textbf{Section~\ref{prel}} will establish the necessary definitions and background knowledge essential for a thorough comprehension of this paper. Following that, \textbf{Section~\ref{relwork}} will provide a comprehensive overview of the related works. Subsequently, in \textbf{Sections~\ref{method}} and\textbf{~\ref{classification}}, we will explain the criteria by which we selected the survey papers and we will dive into the specifics of our developed taxonomy. Later, we will explain our findings in \textbf{Section~\ref{findings}} and open problems in \textbf{Section~\ref{openprob}} respectively. Lastly, we will provide our concluding comments in Section~\ref{conclusion}.

\section{Preliminaries}
\label{prel}
Throughout the paper, we will use the terms \llms{} (\llmsabr{}), Generative AI, AI tools and \apps{} interchangeably. Similarly, we will also assume software developers and users as the same entity. In the following, we provide some short forms and standard definitions used throughout the paper.

\subsection{Short Forms}
\begin{itemize}
    \item LLM - Large Language Model
    \item HCI - Human-Computer Interaction
    \item SE - Software Engineering
    \item ML - Machine Learning
    \item AI - Artificial Intelligence
    \item UML - Unified Modelling Language
\end{itemize}
\subsection{Definition}
\begin{itemize}
    \item Generative AI - A class of AI capable of generating text, images and various other forms of data.
    \item Large Language Model - Specialized Generative AI that are trained on extremely large datasets and can produce various forms of artifacts from natural language prompts.
    \item Polyadic Systems - An intelligent system that can concurrently accommodate the interactions of multiple people.
    \item Mental Model - A mental model represents an individual’s perception of how a particular process works.
    \item Boilerplate Code - Structural code snippets that are necessary for the coding environment but do not contribute towards business logic.
    \item Software Artifact - Any items related to software development, e.g., source code, design document, UML diagrams.
    \item Open Coding - A systematic qualitative approach of grouping data to identify patterns and themes.
    \end{itemize}

\section{Related Works}
\label{relwork}
Researchers have continuously evaluated these \apps{} since their inception to identify performance, best practices, risks, and usability factors. Extensive research highlights both the advantages and drawbacks of this paradigm. Perceived benefits encompass the easy execution of routine tasks, swift generation of boilerplate code, code refactoring, project architecture creation, and unit test generation~\cite{sarkar2022like, imai2022github, ahmad2023towards, yuan2023no}. Some researchers claim that \llmsabr{} offer superior code quality in terms of validity, precision and complexity, while also conserving both time and effort, especially compared to traditional human pair programmers~\cite{yetistiren2022assessing}. However, these advantages are not without setbacks~\cite{nguyen2022empirical}.

It is well-documented that the performance of \llmsabr{} is correlated with their training data quality. In general, the vast majority of the training dataset comes from publicly available software artifacts, and often, these data can be of questionable quality or riddled with vulnerabilities. In addition, \llmsabr{} cannot infer what it is producing. It produces an output that calculates the probability of each token generated next. Thus, it is often affected by bias. If training data contain significantly lower-quality artifacts, it is unlikely that \llmabr{} will produce a good artifact during inference.

Furthermore, the unpredictable nature of \llmsabr{} indicates that they can produce varying outputs for identical prompts~\cite{sajed}. Often, these variations are not limited to semantic structure but also to the overall concept. The privacy implications concerning these models also require attention~\cite{ahmad2023towards}. Developers are now concerned about public exposure to their software system internals. In addition, the generated suggestions often cannot consider the general context of the code base due to the context limit, leading to mistrust by developers~\cite{sorte2015use}.

Moreover, developing \ai{} tools and predicting their behaviors pose significant challenges. Unlike many traditional non-\ai{} systems, these \ai{} systems rely on complex deep learning-based transformer models, which are inherently difficult to assess. In traditional software development, we typically utilize deterministic test cases to evaluate against specific and predictable behaviors. However, given the nondeterministic nature of \llmsabr{}, testing them with such conventional deterministic test cases is a significant challenge and does not align well~\cite{ebert2023generative, yetistiren2022assessing}.

Many of the earlier \apps{} come with a set of fixed interaction features. They limit how a developer can use or interact with the \ai{} tools. These tools often lack the disambiguation and refinement of the query or action specified by a developer~\cite{mcnutt2023design}. To address these usability features, several leading \llm{} distributors have developed a chat-based counterpart such as Copilot X~\cite{copliotx}, Tabnine Chat~\cite{tabninechat} and Cursor~\cite{cursor}. These chat-based systems provide developers more independence in refining the goal and utilizing code context. 

Another key aspect identified in prior studies is understanding the dynamics between human and \ai{} collaborations. There appears to be a diverse range of opinions among researchers and end users on this matter. Subsequent sections will further explain the pros and cons of AI pair programmers in some key software development domains.

\section{Methodology}
\label{method}
To conduct our literature review, we followed the process used in a previous literature review~\cite{zheng2022ux}. This procedure involves four distinct stages: Definition, Search, Selection, and Analysis. A detailed illustration can be found in the figure.

\subsection{Definition}
In this section, we will state how we defined the scope of our study. We made some specific inclusion and exclusion rules for the domain we selected.

\begin{itemize}
    \item We included papers from SE, Generative AI, and HCI domains.
    \item We also considered recent arXiv papers only if they were highly cited or from a reputed author in the relevant domain.
    \item We excluded papers unrelated to developer tools or their usability.
    \item We have considered publications from 2015 and beyond.
\end{itemize}

\subsection{Search}
In order to identify relevant scholarly literature, we curated a list of well-respected conferences and journals within the fields of SE and HCI. Notable venues included ICSE, UIST, CHI, TOSEM, VL/HCC, FSE, ACM Transactions, IEEE software, and MSR. In our preliminary exploration of Google Scholar, we employed keywords such as `llm', `software engineering', `user study', and `hci'. Subsequently, we identified and prioritized the top results from the listed venues. Additionally, we looked into contemporary technology blogs and articles authored by renowned researchers.

The initial pool of collected papers served as a foundation for an iterative snowballing process to uncover further related publications within our domain of interest. Based on the frequency of relevant publications discovered, we prioritized our search focus on specific venues.

\subsection{Selection}
We aimed to select and review publications that most comprehensively represent previous research in our chosen domain. To achieve this objective, we prioritized publications that had received awards and attention within their respective venues. Given that our study investigates the impact on human developers, we preferred publications featuring user studies or related empirical research.

Upon compiling the final list of selected publications, we discarded some of them. It was necessary to make it viable for a single author to comprehensively review them all. Particular attention was paid to maintaining representation from multiple venues.

\subsection{Analysis}
To present our findings with the utmost precision and impartiality, we adopted a systematic and open-minded approach. To achieve this, we used thematic analysis as our methodology. We reviewed each of the selected papers individually. During this process, we identified and carefully documented relevant and intriguing findings. Subsequently, we applied an open-coding approach in a bottom-up fashion, extracting key themes and patterns from the data. Finally, we further refined and validated these findings to obtain concrete results.

Furthermore, we conducted an extensive review of articles, blog posts, and developer surveys that have significantly influenced the \app{} domain. All the curated selections from our review are consolidated in Table~\ref{tab:publications_list} and Table~\ref{tab:reports_list}.

\begin{table*}
    \centering
    \caption{Classification of factors with AI tools influencing software development }
    \label{tab:classificationlist}
    \begin{tblr}{
  colspec={|c|l|l|l|l|l|},
  vlines,
  hlines,
  vspan=even,
  row{1} = {bg=blue, fg=white},
  rowsep=0.2pt
}
S.N.& Category&Sub Category & S.N.& Category&Sub Category\\\hline
         \SetCell[r=7]{}{1}& \SetCell[r=7]{}{\categoryOne{}}&  Web Search & \SetCell[r=4]{}{6}& \SetCell[r=4]{}{\categorySix{}}&Personal Project \\
         &  &  Wiki Sites & & &Small Team \\ 
         &  &  Co-worker & & &Large Team \\ 
         &  &  Documentation & & &Multinational Team \\ 
         &  &  IDE Suggestions & \SetCell[r=2]{}{7}& \SetCell[r=2]{}{\categorySeven{}}&Developer \\  
         &  &  Public \llmabr{} & & &AI Tool \\ 
         &  &  Private \llmabr{} & \SetCell[r=4]{}{8}& \SetCell[r=4]{}{\categoryEight{}}&Newcomer Onboarding \\ 
 \SetCell[r=7]{}{2}&\SetCell[r=7]{}{\categoryTwo{}}&  Requirement Analysis & & &Creativity \\ 
         &  &  Architectural Design & & &Best Practices \\ 
         &  &  Code Generation & & &Knowledge Gain \\ 
 & & Testing & \SetCell[r=5]{}{9}& \SetCell[r=5]{}{\categoryNine{}}&Accuracy \\ 
 & & Code Review & & &Efficiency \\ 
 & & Monitoring & & &Discoverability \\
 & & Maintenance & & &Productivity \\ 
 \SetCell[r=5]{}{3}&\SetCell[r=5]{}{\categoryThree{}}& Inline Level & & &Quality \\
 & & Method Level & \SetCell[r=6]{}{10}& \SetCell[r=6]{}{\categoryTen{}}&Trust \\
 & & Class Level & & &Privacy \\
 & & Module Level & & &Bias \\
 & & Project Level & & &Software License Violation \\
 \SetCell[r=5]{}{4}&\SetCell[r=5]{}{\categoryFour{}}& Computational Notebook & & &Reliability \\
 & & UI Design & & &Training Data Quality \\
 & & Web Development & \SetCell[r=7]{}{11}& \SetCell[r=7]{}{\categoryEleven{}}&Cognitive Load \\
 & & Autonomous Systems & & &Ease of Use \\
 & & Quantum Computing & & &Time to Completion \\
 \SetCell[r=4]{}{5}&\SetCell[r=4]{}{\categoryFive{}}& Computer Science Illiterate & & &Time to Success \\
 & & Novice & & &\\
 & & Intermediate & & &\\
 & & Expert & & &\\
    \end{tblr}
    \vspace{3px}
\end{table*}

\section{Area Classification / Taxonomy}
\label{classification}
Our investigation of the influence of \llmsabr{} on software development encompasses the domains of Software Engineering, Machine Learning, and Human-Computer Interaction. Potential impacts and changes within these spheres are shaped by a multitude of factors. 

To the best of our knowledge, there is no single classification paradigm that covers all of these areas. Recognizing this gap, we introduce a novel evaluation metric that, in our estimation, is not only robust but also invaluable for subsequent research endeavors. \textbf{We came up with a total of 11 categories and 53 subcategories in our classification.} This classification scheme is outlined in Table ~\ref{tab:classificationlist}.

\section{Survey Findings}
\label{findings}
Our investigative study encompassed a review of \totalSurveyPapers{} distinct publications, inclusive of conference proceedings, journal articles, and developer surveys. The collective outcome extends beyond the confines of the papers specifically referenced in Table~\ref{tab:publications_list} and Table~\ref{tab:reports_list}. In addition to these documented findings, we have also integrated our own insights and interpretations. The following discussion will state the findings of our survey with our established classification schema.

\subsection{\categoryOne{}}
Software developers often seek help from numerous sources when they face challenges in their tasks. These challenges primarily revolve around searching, navigating, and interpreting source codes and associated documentation. Expertise in these domains is imperative for activities like feature development and code refactoring, debugging, familiarizing oneself with new technologies, and deployment.

Often, developers move towards resources like search engines such as Google, knowledge sources like Stack Overflow, any related documentation, or IDE suggestions~\cite{sorte2015use}. In an industry setting, reaching out to colleagues for insights is a conventional practice. However, this mode of workflow has been greatly influenced by the advent of \llmsabr{}. Developers are now frequently consulting AI-powered platforms such as ChatGPT, Copilot, and Tabnine, resulting in significant savings in time and effort. Current research highlights an increasing preference for these AI tools, overshadowing traditional IDE suggestions~\cite{sarkar2022like,sofSurvey}. Furthermore, the use of such tools can help manage and preserve legacy codes~\cite{ebert2023generative}.

In light of growing privacy concerns, many developers are inclining towards customized \llmsabr{} solutions or opting for sandbox options such as Ollama~\cite{ollama}. Recent research suggests that a sandboxed environment bolsters security measures~\cite{liu2023wants}. Concurrently, another industrial narrative offers guidelines on architecting enterprise-grade \llmsabr{} systems~\cite{githubTutorial}. The transition from traditional avenues of assistance to AI-assisted tools is noticeable among an increasing number of developers. These \llmabr{} platforms, often trained with vast datasets sourced from the Web, demonstrate capabilities across a wide spectrum of challenges.

\subsection{\categoryTwo{}}
Throughout the software development steps, there has been an extensive application of \llmsabr{}. Historically, the focus was on automating workflows through various tools, limited primarily to specific technology, method, and context~\cite{sorte2015use}. 

\subsubsection{Requirement Analysis}
Traditionally, requirement analysts concentrated on collecting and authenticating client needs. However, with the evolution of AI tools, this process has been enhanced by pinpointing inconsistencies, recognizing overlooked elements, and highlighting key requirement areas. One analysis suggests that such tools aid analysts in drafting more concise user stories~\cite{tikayat2023agile}.

\subsubsection{Architectural Design}
A recent study showcased how ChatGPT collaborated with a novice software architect and successfully crafted a design document~\cite{ahmad2023towards}. However, the results were not evaluated by expert architects to determine whether AI-assisted design is comparable to or exceeds human-driven designs. This domain is known to face challenges due to limited expertise, changing requirements, and accumulated technical debt. The effectiveness of AI-supported design against purely human-generated designs remains a topic of future exploration.

\subsubsection{Source Code Generation} 
The capability of \apps{} is particularly evident when dealing with source code. These tools can synthesize, modify, refactor, and even translate code from one programming language to another~\cite{sarkar2022like}. Developers can prompt their intentions in natural language to obtain the associated code as output.

However, it is not a flawless approach. Developers sometimes struggle with inaccurate outputs and often fail to convey the exact requirements. Research indicates that these tools often produce non-compilable code~\cite{nguyen2022empirical}. Our survey findings suggest this is mainly due to the focus on the local context rather than the project context. As a remedy, future tools should augment broader project context when generating code.

\subsubsection{Testing}
The automation of multiple software testing stages has always been of significant research interest. With \llms{}, often generating test cases from either source code or requirements is now feasible. Our observation shows that these AI tools can generate test cases specifying test domains that a tester might have missed. Often, the resulting tests are comparable to human-made ones and can be quickly adopted by developers~\cite{yuan2023no}. However, such tools have their pitfalls. They might produce compilation errors and unreliable tests, create out-of-scope tests, and generate resource-intensive tests. \textbf{Blind trust in these tools could also lead to less scrutiny of the auto-generated code}~\cite{imai2022github, ebert2023generative}. We identify minimizing compilation errors in test case generation as a necessary future area of improvement.

\subsubsection{Code Review, Monitoring \& Maintenance}
Insights provided by \llmsabr{} can be used to review the code for accuracy, vulnerability, and security violations. However, they can misguide the developers. A deeper understanding of the context can make them more efficient in code maintenance~\cite{liang2023understanding}. There are ongoing efforts to recommend feature additions and initiate pull requests to automatically address issues in open-source settings. We believe that these tools have a considerable journey ahead before complete automation becomes a reality.

\subsection{\categoryThree{}}
Different \apps{} offer code suggestions that apply to various scopes. In general, these suggestions can be categorized into the inline, method, class, module, and project levels.

\subsubsection{Inline}
Tools like Copilot can offer inline code edits. Inline refers to the suggestions provided in the line where a user is holding the cursor. However, users sometimes struggle to invoke this feature, and its performance does not always align with their expectations. Research indicates that certain inline suggestions are not preferred by users, who instead lean towards pop-ups and sidebars as their preferred format~\cite{mcnutt2023design}. A separate industry-supported user study on Visual Studio IntelliCode~\cite{vsintellicode} analyzed the most appreciated form of inline recommendations~\cite{vaithilingam2023towards}. \textbf{The findings revealed that users frequently do not notice the AI assistance available for specific code segments.} We call for further usability studies with more participants to reach a grounded decision.

\subsubsection{Method}
This domain is significantly focused by most \apps{}. Typically, developers input a request in natural language to receive documentation, source code, or unit tests. This area has witnessed substantial research and several benchmark datasets to assess the performance of AI-driven tools. As discussed in earlier sections, a key challenge is the tendency to overlook the broader project context while producing suggestions in this domain~\cite{liang2023understanding}.

\subsubsection{Class, Module \& Project Level}
Initial outlines crafted by \llmsabr{} can serve as a valuable starting point in larger code contexts. They benefit developers by supplying boilerplate and template code, improving the coding process. The results are usually more accurate when the underlying training data contains a similar code at a large frequency. However, this level also faces the highest number of compilation errors. Project or class templates proposed by these tools often contain bugs and incomplete snippets and are prone to library version mismatches. Consequently, \textbf{developers tend to invest additional time deciphering and rectifying the auto-generated code to ensure its functionality.}

\subsection{\categoryFour{}}

\subsubsection{Computational Notebooks}
Computational notebooks present a unique framework tailored mainly for machine learning and data science endeavors. They offer a trial-and-error approach due to the various factors involved. However, the potential integration of \llmsabr{} into this arena remains largely unexplored. User studies have identified key challenges, such as inadequate documentation, hyperparameter fine-tuning, efficient model crafting, dataset preparation, and model debugging. \textbf{Tools designed for this domain often only address experts in data science, ignoring novices}~\cite{wang2022documentation}. This oversight leaves open problems for future work.

\subsubsection{UI Design}
Numerous \llmsabr{} tools aim to assist developers in UI design efforts~\cite{liu2023wants}. Notably, DALL·E~\cite{dalle} has gained traction for its ability to generate images from natural language descriptions. Additionally, AI tools like Galileo AI~\cite{gallilleo} focus exclusively on developing user interfaces. In the traditional way, a considerable time investment is required for interface design. Therefore, more and more artists and designers are using them as a means of better productivity. Several concerns arise regarding the originality of generated images, especially with potential copyright infringements within training datasets. This scenario alarms the creative community, \textbf{prompting suggestions for blockchain-backed AI-generated content to ensure authenticity}~\cite{ebert2023generative}.

\subsubsection{Web Development}
The vast landscape of web development contains the highest concentration of developers. Most of the \llmsabr{} applications find their foundation here, addressing the back-end, front-end, and database angles. Amazon Q~\cite{amazonq} is such a \llmsabr{} integrated within the AWS cloud. They assist developers in tasks such as code translation, integration, boilerplate code creation, and defect resolution. However, existing hurdles such as clarifying ambiguous requirements~\cite{mcnutt2023design}, safeguarding sensitive data like tokens and credentials~\cite{liu2023wants}, and navigating continuously evolving libraries~\cite{oreilly}.

\subsubsection{Autonomous Systems \& Quantum Computing}
Our investigation highlights the promising role of \llmsabr{} in advancing and validating autonomous systems like robots, autonomous medical devices, and self-driving cars~\cite{ebert2023generative}. Additionally, the recent popularity of quantum computing presents opportunities for integration of \llmsabr{}~\cite{serrano2022quantum}. As quantum computers remain a limited entity confined to several tech giants, their broader applications have yet to gain mainstream attention.

\subsection{\categoryFive{}}
\subsubsection{Computer Science Illiterate}
Almost little to no study exists on how we can utilize \llmsabr{} for this domain of users. We identify a significant potential in this domain with ample opportunities. \textbf{Employing \llmsabr{} can serve as an attractive tool for introducing more people to programming, simplifying the learning curve}. The educational sector offers vast future possibilities with \llmsabr{}.

\subsubsection{Novice Users}
Numerous studies highlight the challenges faced by novice developers~\cite{liu2023wants}. Our survey indicates that to fully utilize the capabilities of \llmsabr{}, domain expertise is necessary for better prompting. For instance, a study involving Copilot observed that beginners tend to rely on natural language prompts, while expert developers employ a blend of natural language and code snippets~\cite{sarkar2022like}. However, the exclusive focus on Copilot in this study may limit the scope of its findings in terms of all \apps{}. Addressing the challenges faced by newcomers remains an open area for future exploration.

\subsubsection{Intermediate Users}
Our analysis reveals a common oversight in clearly addressing and identifying intermediate users. Most user studies revolve around novices or experts. We believe that it is crucial to distinctly identify this group given that a majority of developers likely fit this bracket.

\subsubsection{Expert Users}
\textbf{\llmsabr{} work best when used by an expert user}~\cite{sarkar2022like, mcnutt2023design}. By expert, we indicate users who have an in-depth understanding of the query domain and \llmabr{} internal working mechanism. This profound understanding enables them to craft clearer prompts that resonate better with \llmsabr{}. A study focusing on expert developers determined that such users are generally more forgiving of generated errors\cite{mcnutt2023design}. Expert users often ask for faster responses and ease of interactions with different shortcuts. \textbf{They find the context switch to utilize \llmsabr{} very distracting. }

\subsection{\categorySix{}}
\subsubsection{Individual Project} 
Often, a developer undertakes a hobby project or aims to create tools to streamline daily tasks. \llmsabr{} can assist in rapidly locating similar projects and understanding new technologies. However, developers may encounter challenges like receiving suggestions for obsolete code, enduring multiple attempts to get a functional prototype, or even giving up.

\subsubsection{Small \& Large Teams}
In team settings, developers struggle with issues like interpreting legacy code, onboarding new team members, and understanding large codebases. Even with available training opportunities and documentation in larger organizations, these activities can be challenging. They sometimes lead to unnecessary pain points, frequent context changes, and possible burnout.

There are opportunities with \llmsabr{} to simplify these processes. AI tools capable of handling vast contexts will prove invaluable in maintaining consistent coding practices across the team~\cite{zheng2022ux}. Larger teams can use custom AI tools for the specific needs and activity styles of the teams~\cite{wang2023investigating}. However, such customization would require an internal expert to fine-tune the \llmabr{} models using proprietary data. \textbf{An alternative could be to use a universally modifiable AI tool that adjusts by learning from code repositories and team dynamics.} However, this can lead to privacy and ethical concerns regarding developer activity monitoring.

\subsubsection{Multinational Teams} 
Teams that span multiple nations present a unique set of challenges, given linguistic, cultural, and geographical variations. Due to these differences, face-to-face discussions or dropping by a coworker's desk to seek assistance are not feasible choices. In such settings, \llmsabr{} can act as an agent for recognizing and accommodating these diversities, guiding developers toward optimal decisions. Past studies on polyadic interactions have proposed that AI tools might play a role in managing and moderating sentiments in communication~\cite{zheng2022ux}.

\subsection{\categorySeven{}}
\subsubsection{Developer}
A mental model represents an individual's understanding or perception of how a particular process works. Usability studies often provide great importance in keeping the working model similar to the user's mental model. Numerous prior studies have analyzed the mental models of software developers~\cite{mcnutt2023design}. Common challenges include search, navigation, and finding similar resources. It is crucial that AI agents understand the way a developer's mind works and should accommodate a working model that connects with the developer's perceived mental model.

\subsubsection{AI} 
Various prior studies try to better understand how deep neural networks in a language model work~\cite {ebert2023generative}. \textbf{Developers must understand the mental model of AI tools so that they can modify their commands and interactions to achieve the best outcome}. For example, how the underlying tokenization, embedddings, and inferencing works. This remains an open problem of how to better make a mental model \llms{} understandable to developers.

\subsection{\categoryEight{}}
\subsubsection{Newcomer Onboarding} There is a pressing challenge in the realm of open-source projects. Unlike traditional industry settings, these projects often lack comprehensive documentation and essential guidelines for newcomers. Ongoing research is trying to address these obstacles. Furthermore, in many industrial scenarios, onboarding new developers through conventional methods is expensive. \textbf{Often, newcomers are assigned to existing codebases without any systematic training.} Addressing these challenges could serve as a promising avenue for future research.

\subsubsection{Creativity} 
\llmsabr{} holds the potential to remove barriers to creative solutions~\cite{ebert2023generative, bird2023taking}. The developer's exploration and acceleration modes can be enhanced with such creative outcomes~\cite{barke2023grounded}. Prioritizing usability with \llmsabr{} is imperative to eliminate cognitive barriers that hinder creativity.

\subsubsection{Best Practices} 
\textbf{Many difficulties arise from the lack of adherence to recognized best practices, including the use of design patterns, defensive coding, and the avoidance of unit testing}~\cite{wang2022documentation, bird2023taking}. Current literature suggests that \llmsabr{} possesses the capability to pinpoint edge cases often overlooked by developers~\cite{liang2023understanding,bird2023taking}. However, these AI models, due to their nondeterministic nature, occasionally produce false positives. Moreover, the commitment to best practices is frequently neglected within the domains of Machine Learning and Data Science.

\subsubsection{Knowledge Gain} 
The potential of Intelligent Assistants to promote knowledge enhancement is of interest to academic researchers and industry professionals~\cite{canbek2016track}. Through the use of \llmsabr{}, people can acquire new insights, including learning new programming languages more efficiently. Our survey also confirmed these assertions. Khan Academy~\cite{khanacademy} has developed Khanmigo~\cite{khanmigo}, an \llmabr{} based learning platform for enhanced learning. \textbf{We also expect a change in educational platforms related to software development}.

\subsection{\categoryNine{}}
In this subsection, we present the positive outcomes identified through our survey. We explain the most occurring beneficial factors observed during our study.

\subsubsection{Accuracy} 
Finding an accurate solution to a problem is an essential aspect of software development. Existing literature confirms that \apps{} can improve task precision, though they often overlook the element of non-determinism~\cite{nguyen2022empirical, yetistiren2022assessing}. While \apps{} may not consistently deliver perfect solutions, they demonstrate usefulness in this domain. Future research could concentrate on stabilizing accuracy in response to specific prompts across multiple iterations.

\subsubsection{Efficiency} 
Working with \apps{} can improve work efficiency. Users transitioning into the acceleration mode exhibit elevated capabilities. For this enhancement, they are often termed as an advanced auto-completion tool~\cite{sarkar2022like, liang2023understanding}.

\subsubsection{Discoverability}
AI tools can demonstrate assistance in discovering problems and bugs or act as a starting point by proposing potential locations of problems. Developers can utilize this tool to summarize the working domain.

\subsubsection{Productivity} 
\textbf{A significant uplift in overall productivity is confirmed by numerous studies}~\cite{liang2023understanding, jain2022jigsaw}. A particular investigation used the volume of code added and omitted as a measure of productivity and quality~\cite{imai2022github}. The research determined that Copilot usage catalyzed a boost in productivity. Further automation in the realms of debugging and disambiguation will enhance future tools~\cite{wang2023investigating}.

\subsubsection{Quality} 
Evaluating the quality of software artifacts does not follow a universal standard. Diverse methodologies exist to assess various aspects of software quality. While a study reports favorable outcomes concerning quality enhancement~\cite{yetistiren2022assessing}, we perceive a gap in selected metrics. One specific study demonstrates that the quality can be affected with the use of Copilot~\cite{imai2022github}.

\subsection{\categoryTen{}}
In the following, we will discuss some key concerns with \apps{} observed in our survey.

\subsubsection{Trust} 
Trust presents a significant barrier in using \apps{}\cite{wang2023investigating}. \textbf{Developers often do not trust AI tools to generate a totally meaningful solution}. The distrust stems from the fact that AI tools may not consistently produce meaningful output. Misinformation, hallucinations, and invalid claims are the responsible factors for reducing trust. Furthermore, their performance tends to deteriorate in large projects, often misinterpreting the existing context\cite{liang2023understanding}.

\subsubsection{Privacy} 
Within industry spheres, developers often express concerns about the privacy used by \apps{}~\cite{ahmad2023towards, ebert2023generative, zheng2022ux, liu2023wants}. For example, uploading proprietary code to a language model could inadvertently disclose code snippets to other users due to the reinforcement learning mechanisms of these models. \textbf{This could potentially expose secret keys and can be exploited by malicious entities for advanced cyber attacks.}

\subsubsection{Bias} 
Bias is a well-known problem with machine learning models~\cite{ahmad2023towards, sarkar2022like}. This arises primarily from the imbalance of the training data. \textbf{These tools sometimes manifest discrimination related to age, gender, race, and origin}. While supplementing and balancing training data serves as a potential mitigation strategy, it becomes increasingly untenable with larger datasets utilized in training Generative AI models.

\subsubsection{Software License Violation} 
Our survey has found that the training data used for code-synthesizing AI indiscriminately encompass all public codebases without consideration of their open-source licenses~\cite{liang2023understanding, bird2023taking}. \textbf{An unsuspecting developer using \apps{} might end up violating these licenses by accepting the suggested code.} We identify this as a domain that requires immediate attention.

\subsubsection{Reliability} 
Prior studies found these tools unreliable due to their inherent nature~\cite{ebert2023generative, jain2022jigsaw, ahmad2023towards}. Since machine learning works with probability at its core, the results often lack consistency and reliability. With the same prompt, developers may get different outputs at different times~\cite{sajed}. Besides, crafting an effective prompt also presents challenges~\cite{sarkar2022like}. One of the ways to address this is to show the user multiple suggestions in the probability domain. But that comes with many usability challenges, leading to more context load and effort and possibly cluttering the display.

\subsubsection{Training Data Quality} 
Despite the regular enhancement of \llms{} by increasing the number of parameters, concerns remain about the quality of the training data. For example, not all public online resources provide correct solutions. Even accepted answers on platforms like Stack Overflow might not address all user needs and contexts. Moreover, these probabilistic models lack semantic understanding, merely predicting the succeeding token based on associated probability.

\subsection{\categoryEleven{}}
The design of new tools frequently overlooks cognitive aspects, leading to usability challenges and unexpected complications~\cite{zheng2022ux}. This subsection discusses the cognitive issues that arise in the context of \apps{}.

\subsubsection{Cognitive Load} 
The primary intent of these language models is to facilitate developers to carry out their tasks effortlessly. An essential component of this is minimizing cognitive burden during interactions. However, \textbf{our survey findings suggest a mixed impact using these tools}~\cite{liu2023wants}. A user study spanning multiple programming languages has observed reduced code complexity with code produced by these models~\cite{nguyen2022empirical}. However, several external factors can adversely affect this area. We observed an oversight of the cognitive burden when evaluating performance outcomes~\cite{imai2022github, yetistiren2022assessing}.

\subsubsection{Ease of Use} 
\textbf{Empirical evidence suggests that users often find these tools perplexing and not easy to navigate}~\cite{liang2023understanding}. The recommendations provided by these tools can occasionally mislead the developer, presenting multiple challenges, including issues with execution history, tool customization, and fine-tuning~\cite{mcnutt2023design, wang2023investigating}.

\subsubsection{Time to Completion \& Success} 
Efficient tools can potentially reduce the time span required to realize a task, encompassing the time to select, implement, and subsequently validate a solution against user specifications. In the industrial realm, time is of great significance and has profound implications for organizational success. Recent research indicates that the use of \apps{} can positively influence this dimension~\cite{liang2023understanding}.

\begin{figure}[ht]
    \centering
    \includegraphics[width=\linewidth]{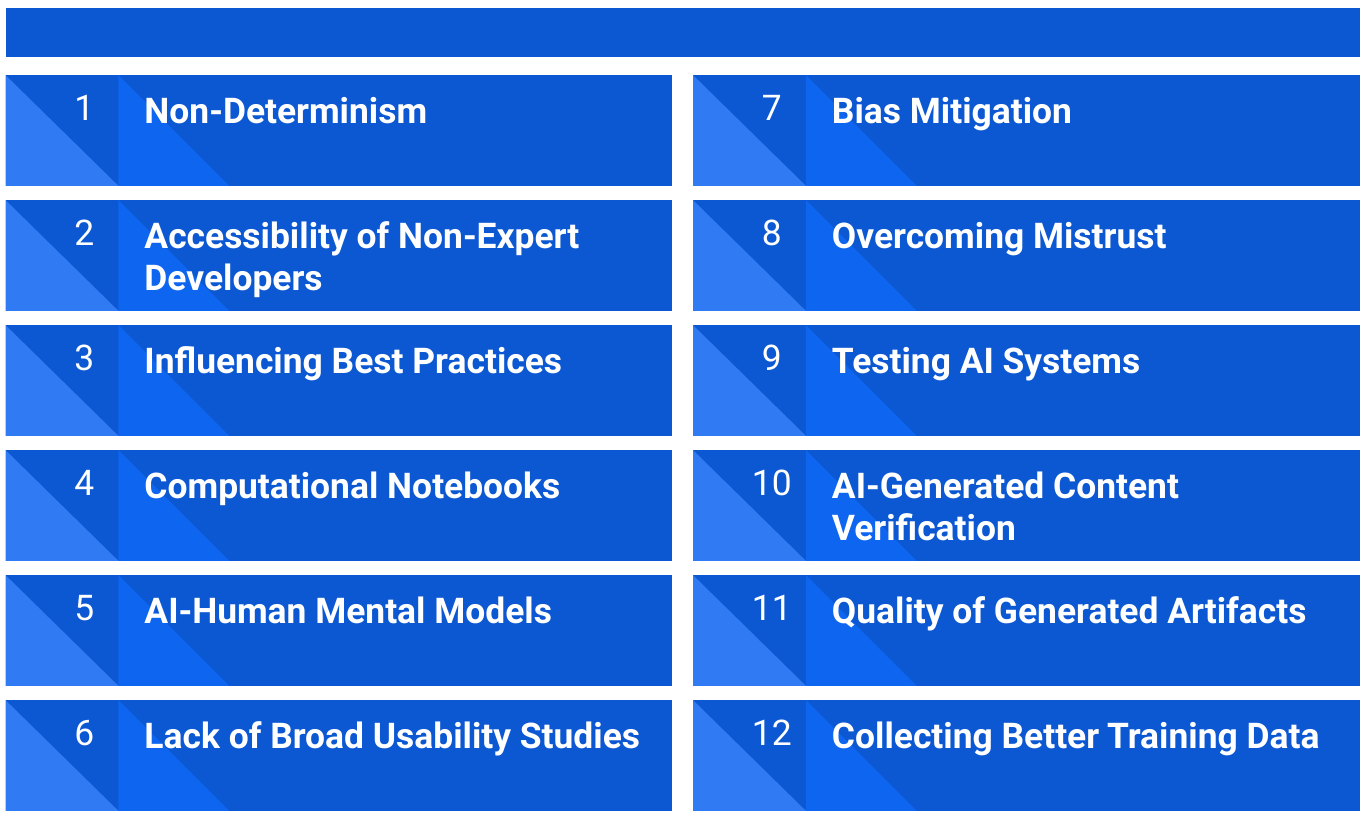}
    \caption{Identified Open Problems}
    \label{fig:open-problems}
\end{figure}

\section{Open Problems}
\label{openprob}
Our survey analysis has revealed several issues that require future studies. Figure \ref{fig:open-problems} shows a curated list of \textbf{12 identified open problems }in the software development domain with \apps{}.

\subsection{Non-Determinism}
The non-deterministic nature of \apps{} is a major concern, leading to numerous issues. This unpredictability lowers user confidence and tool accuracy. Addressing and minimizing this inconsistency is crucial for future research.

\subsection{Accessibility of Non-Expert Developers}
Our analysis indicates that \apps{} are designed considering expert developers. There is an evident need to make these tools more approachable and usable for all user levels. Moreover, these tools could serve as invaluable learning resources for beginners and those without a computer science foundation.

\subsection{Influencing Best practices}
While numerous studies outline best practices across domains, none enforce these within \apps{}. As reliance on these tools increases globally, it becomes essential for them to embed and advocate for these best practices. We believe this approach could simplify developer workflows and review processes.

\subsection{Computational Notebook}
The computational notebook domain faces greater challenges in accommodating novices, maintaining documentation, and developing fine-tuned AI models. Since they are different from traditional programming paradigms, they require special attention. Existing research in this domain with \apps{}, only addresses a tiny portion of the problems.

\subsection{AI-Human Mental Models}
Understanding the factors that drive AI responses remains complex. Most current usability studies focus on human mental models. To enhance AI-human collaboration, research must consider harmonizing both AI and human mental constructs.

\subsection{Lack of Broad Usability Studies}
Most user studies are observational and often feature a limited participant pool, usually 15 to 30. Moreover, they often consist of only graduate students. Popular forms of large user studies consist of survey responses. However, they are often unreliable since they are self-reported. This poses an open challenge to perfectly identify the sentiments of a diverse population.

\subsection{Bias Mitigation}
Combatting biases in AI remains an ongoing issue. Current approaches used by smaller machine learning models are not suitable for large models like \apps{}. Given the global adoption of these tools, we recommend prioritizing bias mitigation.

\subsection{Overcoming Mistrust}
Mistrust is a key concern observed among the developers. Our survey found evidence that for larger projects, the responses by \apps{} are unreliable. Even for smaller tasks, the generated responses might vary significantly. Therefore, addressing mistrust is key to more AI acceptance.

\subsection{Testing AI systems}
In Table~\ref{tab:publications_list}, we have reported several benchmark datasets to test the performance of the \apps{}. However, we find these datasets lacking since they address only smaller tasks and do not test the larger tasks or repetitive user interactions. Furthermore, setting up a suitable development environment contributes many factors when working with these AI tools. Therefore, we recommend a more comprehensive testing approach.

\subsection{AI-Generated Content Verification}
Creative professionals are now challenged with the verifiability of AI-generated content, which is specifically problematic for UI designers. Concurrent copyright concerns amplify these challenges. One of our survey papers suggests using blockchain-based end-to-end verification. However, there are no comprehensive studies to verify its suitability. We must come up with more reliable mechanisms to verify AI-generated content.

\subsection{Quality of Generated Artifacts}
Contents generated with \apps{} are often confusing and contain compilation errors. Several studies in our survey have affirmed this phenomenon. Besides, these tools need to produce artifacts with better precision. This is a challenging domain since expecting good-quality responses from questionable-quality training datasets is not ideal.

\subsection{Collecting Better Training Data}
Most of the training data for \llms{} are sourced from the public website domains. The quality and validity of these data are inherently questionable. However, gathering such a vast amount of synthetic data is challenging. Researchers need to address this issue with more innovative paradigms.

\section{Conclusion}
\label{conclusion}
\llms{} have dramatically transformed software development, marking a significant shift in how developers engage with their work. These models serve as tools and intelligent companions in the development process, assisting in tasks ranging from code creation to generating innovative ideas and detecting potential issues. Despite numerous challenges, software engineers have adopted them extensively in their routine activities. 

Their swift adoption highlights their effectiveness in boosting productivity. It is evident from our study that \llmsabr{} will continue to be crucial in software development in the coming years. A collaborative partnership between developers and \llmsabr{} will be essential for ongoing progress and innovation.

\bibliographystyle{ieeetr}
\bibliography{main}

\end{document}